\documentclass{jnmp}
\usepackage{amsmath}

\JNMPnumberwithin{equation}{section}

\theoremstyle{definition}

\begin{document}
\renewcommand{\evenhead}{B. A. Kupershmidt}
\renewcommand{\oddhead}{KdV6:  An Integrable System}
\thispagestyle{empty}

\Name{$\mathbf{KdV6}$: An Integrable System }

\label{firstpage}

\Author{Boris A. Kupershmidt~$\dag$}
\Address{$^\dag$ The University of Tennessee Space Institute \\
~~Tullahoma, TN  37388, USA \\
~~E-mail:  bkupersh@utsi.edu\\[10pt]}


\begin{abstract}
\noindent
$K^2 S^2 T \ [5]$ recently derived a new 6$^{th}$-order wave equation $KdV6$: \ $(\partial^2_x + 8u_x \partial_x + 
4u_{xx})(u_t + u_{xxx} + 6u_x^2) = 0$, found a linear problem and an auto-B${\ddot{\rm{a}}}$ckclund transformation for 
it, and conjectured its integrability in the usual sense.  We prove this conjecture by constructing an infinite 
commuting hierarchy $KdV_n6$ with a common infinite set of conserved densities.  A general construction is presented 
applicable to any bi-Hamiltonian system (such as all standard Lax equations, continuous and discrete) providing a 
nonholonomic perturbation of it.  This perturbation is conjectured to preserve integrability.  That conjecture is 
verified in a few representative cases:  the classical long-wave equations, the Toda lattice (both continuous and 
discrete), and the Euler top.
\end{abstract} 

\section{Introduction}

The theory of differential and difference Lax equations has been well understood by the middle of the 1980s, and no 
surprises have disturbed the contented tranquility of the subject ever since.  Until now.

Recently, the 5 authors of [5] subjected to the Painlev${\acute{\rm{e}}}$ analysis the 6$^{th}$-order nonlinear wave 
equation
\begin{gather}
u_{xxxxxx} + au_x u_{xxxx} + bu_{xx} u_{xxx} + cu^2_x u_{xx } + \notag \\
+ du_{tt} + eu_{xxxt} + f u_x u_{xt} + gu_t u_{xx} = 0, \tag{1.1} \
\end{gather}
where $a, b, c, d, e, f$ and $g$ are arbitrary parameters, and they have found 4 cases that pass the 
Painlev${\acute{\rm{e}}}$ test.  Three of these were previously known, but the 4$^{th}$ one turned out to be new (eqn 
(5) in [5]): 
\begin{equation}
(\partial^2_x + 8u_x \partial_x + 4u_{xx})(u_t + u_{xxx} + 6u_x^2) = 0 . \tag{1.2}
\end{equation} 
This equation, as it stands, doesn't belong to any recognizable theory.  The $K^2 S^2 T$ convert it, in the variables 
$v = u_x, \ w = u_t + u_{xxx} + 6u^2_x$, into (eqn (12) in [5]):
\begin{equation}
v_t + v_{xxx} + 12vv_x - w_x = 0, \ w_{xxx} + 8 vw_x + 4wv_x = 0. \tag{1.3}
\end{equation}
If integrable, this is a truly remarkable system:  since $\{ w = 0\}$ leaves only the unperturbed $KdV$ itself, and 
the constrain on $w$ is {\it{differential}}, what we have here is a {\it{nonholonomic}} deformation of the $KdV \ (= 
KdV_2)$ equation.  The 5 authors of [5] found a linear problem and an auto-B${\ddot{\rm{a}}}$cklund transformation for 
their equation, but reported that they were unable to find higher symmetries on available computers, and asked if 
higher conserved densities and a Hamiltonian formalism exist for their equation.

All  of these queries are resolved affirmatively below.

We now proceed to the general construction of nonholonomic perturbations of bi-Hamiltonian systems.  Rescaling $v$ and 
$t$ in the equation (1.3), we get:
\begin{equation}
u_t = 6uu_x + u_{xxx} - w_x, \ \ w_{xxx} + 4 uw_x + 2 u_x w = 0. \tag{1.4}
\end{equation}
This can be converted into
\begin{equation}
u_t = B^1 \bigg(\frac{\delta H_{n+1}}{\delta u}\bigg) - B^1 (w) = \ \ B^2 \bigg(\frac{\delta H_n}{\delta u}\bigg) - 
B^1 (w), \ \ B^2 (w) = 0, \tag{1.5}
\end{equation}
where
\begin{equation}
B^1 = \partial = \partial_x, \ \ \ B^2 = \partial^3 + 2 (u \partial + \partial u) \tag{1.6}
\end{equation}
are the two standard Hamiltonian operators of the $KdV$ hierarchy, $n =2$, and
\begin{equation}
H_1 = u, \ H_2 = u^2/2, \ H_3 = u^3/3 - u_x^2/2, ... \tag{1.7}
\end{equation}
are the conserved densities.
\\
\\
And {\it{that's it.}}  The ansatz (1.5) provides a nonholonomic deformation of {\it{any}} bi-Hamiltonian system.  The 
question, naturally, is whether this ansatz is reasonable or an absurd phantasy. My answer is two-fold:  (A) It is 
reasonable; (B) It is difficult, if not impossible, to prove integrability in general.  The arguments are as follows.  
\\
\\
(A) \ {\it{Each}} system (1.5) has an infinite sequence of $H_m$'s as its conserved densities:
\begin{gather}
\frac{dH_m}{dt} \sim \frac{\delta H_m}{\delta u} \frac{\partial u}{\partial t} = \frac{\delta H_m}{\delta u} [B^2 ( 
\frac{\delta H_n}{\delta u}) - B^1 (w)] \ \sim B^1 (\frac{\delta H_m}{\delta u})w = \notag \\
= B^2 (\frac{\delta H_{m-1}}{\delta u}) w \sim - \frac{\delta H_{m-1}}{\delta u} B^2 (w) =0, \tag{1.8}\
\end{gather} 
where, as usual, $a \sim b$ means:  $(a-b) \in Im \partial$ (a ``trivial Lagrangian'').
\\
\\
(B) \ The above calculation is about the only one that can be reliably performed in the $\{u; w\}$-picture, because 
the constraint $B^2 (w) =0$ is {\it{nonholonomic.}}  Thus, if we proceed to develop the variational calculus in the 
$\{u; w\}$-variables, we would be blocked, because the calculus works {\it{only}} when the factor $\Omega^1/\partial 
(\Omega^1)$ is a {\it{free}} module, where $\Omega^1$ is the module of differential 1-forms  (see [8]).  Thus, the 
question of integrability:  whether the flows (1.5) still commute between themselves, can not be answered in general 
with the modern tools.  It {\it{can}} be answered for the $KdV$ case (and I believe for all the standard differential 
Lax equations) through a subterfuge.  To get a hint on how to proceed, we start in the next Section with the classical 
long-wave system
\begin{equation}
u_t = h_x + uu_x, \ \ h_t = (uh)_x. \notag \
\end{equation}
Section 3 is devoted to the $KdV_n6$ hierarchy (1.5,6) itself.  Section 4 treats the Toda lattice and its continuous 
limit.  The last section considers the classical Euler top.
\\
{\bf{Remark 1.9.}}  The term {\it{nonholonomic}} is of a recent vintage, and seems to have been invented by Hertz, see 
[3].

\section{The Classical Long-Wave Equations}

The classical long wave system is bi-Hamiltonian (in fact, 3-Hamiltonian) $[11], [6], [8]$:  
\begin{gather}
{u \choose h}_t = \partial {h+u^2/2\choose uh} = {0 \ \partial \choose \partial \ 0} {\delta/\delta u \choose 
\delta/\delta h}\bigg(\frac{u^2 h+h^2}{2}\bigg) = \tag{2.1$a$} \\
= \left(\begin{array}{cc}
2 \partial & \partial u \\ 
 u \partial &  h \partial + \partial h \end{array} \right)  {\delta/\delta u \choose \delta/\delta h} 
\bigg(\frac{uh}{2}\bigg). \tag{2.1$b$} \
\end{gather}
Thus, its perturbation (1.5) is:
\begin{gather}
u_t = (h+ u^2/2)_x - w_{2,x}, \ \ h_t = (uh)_x - w_{1, x}, \tag{2.2$a$} \\
(2w_1 + uw_2)_x = 0, \ \ u w_{1,x} + (h \partial + \partial h)(w_2) = 0. \tag{2.2$b$} \
\end{gather}
The first of the constrains in (2.2$b$) is resolvable, but the second one is not, and we seem to be stuck.  The help 
comes from the missing from (2.1) (gravity) parameter $g$ [1], rescaled away for mathematical simplicity (which was 
immaterial in the holonomic framework, but is fatal in the nonholonomic case):
\begin{gather}
{u \choose h}_t = \partial {gh + u^2/2 \choose uh} = {0 \ \partial \choose \partial \ 0} {\partial /\partial u \choose 
\partial /\partial h} \bigg(\frac{u^2 h + gh^2}{2}\bigg) = \tag{2.3$a$} \\
= \left(\begin{array}{cc}
2g \partial & \partial u \\ 
 u \partial &  h \partial + \partial h \end{array} \right)
 {\delta/\delta u \choose \delta/\delta h} \bigg(\frac{uh}{2}\bigg), \tag{2.3.$b$} \
\end{gather}
so that now
\begin{gather}
u_t = (gh + u^2/2 - w_2)_x, \ \ h_t = (uh - w_1)_x, \tag{2.4$a$} \\
(2gw_1 + uw_2)_x = 0, \ \ uw_{1,x} + (h \partial + \partial h)(w_2) = 0. \tag{2.4$b$} \
\end{gather}

The constraint (2.4$b$) is {\it{resolvable}} as a regular series in $g$.  This can be seen as follows.  The first eqn 
in (2.4$b$) yields:  $2gw_1 + uw_2$ = function of $t$ and $\epsilon$ only, and we rescale that function into 1:
\begin{gather}
2gw_1 + uw_2 = 1 \Rightarrow \tag{2.5}\\
w_2 = (1 - 2 gw_1)/u \Rightarrow \tag{2.6}\\
w_{1,x} = - \frac{1}{u} (h \partial + \partial h) \frac{1}{u} (1 - 2gw_1) = \bigg( - \frac{h}{u^2}\bigg)_x + 2g 
\bigg(\frac{h}{u^2} \partial + \partial \frac{h}{u^2}\bigg) (w_1). \tag{2.7} \
\end{gather}
Set now
\begin{equation}
w_1 = \sum^\infty_{k=0} g^k z_k. \tag{2.8$a$}\
\end{equation}
Then
\begin{gather}
z_0 = - h/u^2, \ z_{k+1} \ = 2 (\frac{h}{u^2} \partial + \partial \frac{h}{u^2})(z_k) \Rightarrow \tag{2.8$b$} \\
w_1 = - \sum^\infty_{k=0} {2k+1 \choose k} g^k (h/u^2)^{k+1} = \frac{1}{2g} [1 - (1 - 4g \frac{h}{u^2})^{-1/2}] 
\Rightarrow \tag{2.9$a$} \\
w_2 = \frac{1}{u}(1 - 4g \frac{h}{u^2})^{-1/2} \ \Rightarrow \tag{2.9$b$} \\
{w_1 \choose w_2} = {\delta / \delta u \choose \delta / \delta h} (G), \ G = \frac{u - \sqrt{u^2 - 4gh}}{2g} . 
\tag{2.10}\
\end{gather}
But $G$ commutes with all the $H_n$'s, because
\begin{equation}
G_{hh} / G_{uu} = 2g/h = H_{n, hh} / H_{n, uu}, \ \ \forall n \in \large{\mathbb{Z}}_{\geq 2}. \tag{2.11}
\end{equation}
Thus, all the flows (1.5) commute also.

The workable approach to our general problem hence is this:
\\ 
$(\hat A)$ Rescale the variables $u$ in (1.5) in such a way that the nonholonomic constrain $B^2(w) =0$ becomes 
resolvable, hopefully in the form
\begin{equation}
w = \delta G/\delta u. \tag{2.11}
\end{equation}
$(\hat B)$ If then
\begin{equation}
\{ G, H_n\} \sim 0, \ \ \forall n, \tag{2.12}
\end{equation}
then all the flows (1.5) commute between themselves. (See the end of Section 4 for more on this.)

Let's see now how this approach works for the $KdV_n6$ case.
\\
{\bf{Remark 2.13.}}  Since the  long-wave system is {\it{three-Hamiltonian}}, the nonholonomic construction applies 
not only to the pair $(B^1, B^2)$, but also to the pair $(B^2, B^3)$ of the corresponding Hamiltonian structures.  
It's not clear how these two different perturbations are related.
\\
{\bf{Remark 2.14.}}  $N$-component systems of hydrodynamical type (= \ 0-dispersion) are trivial for $N < 3$, but 
their honholonomic perturbations are no longer so.

\section{\mbox{\boldmath$KdV_n6$}}

We rescale $\partial_t$ and $\partial_x$ by $\epsilon$.  The $KdV_n6$ (1.4-7) becomes (1.5), now with
\begin{gather}
B^1 = \partial, \ B^2 = \epsilon^2 \partial^3 + 2(u \partial + \partial u), \tag{3.1}\\
B^2 (w) = \epsilon^2 w_{xx} + 2 (u \partial + \partial u) (w) = 0. \tag{3.2} \
\end{gather} 
To solve (3.2), set
\begin{equation}
w = \sum^\infty_{k=0} \epsilon^k w_k. \tag{3.3}
\end{equation}
We get:
\begin{equation}
w_0 = u^{-1/2}, \ w_1 = ..., \tag{3.3}
\end{equation}
and in fact
\begin{equation}
w = \frac{\delta G}{\delta u}, \ \ G = \sum^\infty_{s=0} u^{1/2-s} {1/2 \choose s} p_s, \ \ \{ G, H_n \} \sim 0, \ \ 
\forall n, \tag{3.4}
\end{equation}
where $p_s$ are certain differential polynomials from the differential algebra $\large{\mathbb{Q}} [u^{(1)}, u^{(2)}, 
... ][\epsilon]$.  The proof is, unfortunately, rather long, and I omit it ([10]).  I believe that similar rescaling 
works for the general differential Lax (= Gel'fand-Dickey) hierarchy, with the Lax operator
\begin{equation}
{\cal{L}} = u + \sum^N_{i=1} u_i (\epsilon \partial)^i, \ \ \ u_N = 1, \ \ u_{N-1} = 0, \ \ N \in 
\large{\mathbb{Z}}_{\geq 3}, \tag{3.5}
\end{equation}
but I haven't proved it.  (The method was explained, for the case of the Burgers hierarchy, in my talk at the AMS 
meeting at Williams College in the fall of 2001.)
\\
{\bf{Remark 3.6.}}  The situation becomes much more complicated when one passes to the {\it{modified}} Lax equations.  
For the KdV6 case, with the Miura map  $u= \epsilon v_x - v^2$, one gets:
\begin{gather}
v_t = \epsilon^2 v_{xxx} - 6v^2 v_x + p, \tag{3.7$a$}\\
(\epsilon \partial - 2 v)(p) = \partial (w), \ (\epsilon \partial - 2v) \partial (\epsilon \partial + 2v) (w) = 0, 
\tag{3.7$b$} \
\end{gather}
so that one has a {\it{pair}} of nonholonomic constrains attached to {\it{one}} scalar field $v$.

\section{The Toda Lattice}

The Toda lattice is a classical mechanical system with the Hamtiltonian
\begin{equation}
H = \sum_n (\frac{p_n^2}{2} + e^{q_{n+1}- q_{n}}). \tag{4.1} 
\end{equation}
In the variables
\begin{equation}
a_n = p_n, \ b_n = exp (q_{n+1} - q_n), \tag{4.2}
\end{equation}
the motion equations become:
\begin{equation}
{a \choose b}_t = \notag 
\end{equation}
\begin{equation}
{(1 - \bigtriangleup^{-1}) (b) \choose b ( \bigtriangleup -1) (a)} = 
 \left(\begin{array}{cc}
 0 & (1 - \bigtriangleup^{-1}) b\\
 b (\bigtriangleup -1) & 0 
 \end{array} \right) {a \choose 1} = 
 \left(\begin{array}{cc}
 0 & (1 - \bigtriangleup^{-1}) b\\
 b (\bigtriangleup -1) & 0 
 \end{array} \right)
 {\delta /\delta a \choose \delta /\delta b} \ (\frac{a}{2}^2 + b) = \notag \
 \end{equation}
 \begin{equation}
 = \left(\begin{array}{cc}
 b \bigtriangleup - \bigtriangleup^{-1}b & a (1 - \bigtriangleup^{-1}) b\\
 b (\bigtriangleup -1)a & b (\bigtriangleup - \bigtriangleup^{-1})b 
 \end{array} \right)
{1 \choose 0}
= \left(\begin{array}{cc}
 b \bigtriangleup - \bigtriangleup^{-1}b & a (1 - \bigtriangleup^{-1}) b\\
 b (\bigtriangleup -1)a & b (\bigtriangleup - \bigtriangleup^{-1})b 
 \end{array} \right) {\delta /\delta a \choose \delta /\delta b} (a), \tag{4.3} \
 \end{equation}
 where $\bigtriangleup$ is the shift operator:  $(\bigtriangleup f) (n) = f (n+1)$, all the equalities are understood 
as between functions of $n \in \mathbb{Z}$ (or $\mathbb{Z}/ N \mathbb{Z}$), and (4.3) shows the first two (out of 
three) Hamiltonian structures, in the $\{a; b\}$-variables, of the Toda lattice (see [7]).

The nonholonomic deformation ansatz (1.5) produces:
\begin{gather}
a_t = (1 - \bigtriangleup^{-1}) (b - b w_2), \ b_t = b (\bigtriangleup -1)(a - w_1), \tag{4.4} \\
(b \bigtriangleup - \bigtriangleup^{-1} b) (w_1) + a (1 - \bigtriangleup^{-1}) (b w_2) = 0, \tag{4.5$a$} \\
b (\bigtriangleup -1) (aw_1 + (1 + \bigtriangleup^{-1})( b w_2)) = 0. \tag{4.5$b$} \
\end{gather} 
The nonholonomic constrain (4.5), as it stands, is unresolvable.  To proceed, we first rescale $b$ into $\epsilon b$ 
and then look for solutions regular in $\epsilon$.  We get:
\begin{gather}
(b \bigtriangleup - \bigtriangleup^{-1}b) (w_1) + a(1 - \bigtriangleup^{-1}) (w) = 0, \tag{4.6$a$} \\
b (\bigtriangleup -1) (aw_1 + \epsilon (1 + \bigtriangleup^{-1}) (w)) = 0, \ \ \ w: = bw_2. \tag{4.6$b$} \
\end{gather}
(4.6$b$) implies that $aw_1 + \epsilon (1 + \bigtriangleup^{-1})(w)$ = function of $t$ and $\epsilon $ only, and we 
rescale it into 1:
\begin{gather}
aw_1 + \epsilon (1 + \bigtriangleup^{-1}) (w) = 1 \Rightarrow \tag{4.7} \\
w_1 = \frac{1}{a} [1 - \epsilon (1 + \bigtriangleup^{-1}) (w)] \Rightarrow \notag \\
- (1 - \bigtriangleup^{-1}) (w) = \frac{1}{a} (b \bigtriangleup - \bigtriangleup^{-1} b) \frac{1}{a} [1 - \epsilon (1 
+ \bigtriangleup^{-1}) (w)] = \notag \\
= \bigg(\frac{b}{aa^{(1)}} \bigtriangleup - \bigtriangleup^{-1} \frac{b}{aa^{(1)}}\bigg) [1 - \epsilon (1 + 
\bigtriangleup^{-1}) (w)] = (1 - \bigtriangleup^{-1})(c) + \notag \\
- \epsilon (c \bigtriangleup - \bigtriangleup^{-1}) (1 + \bigtriangleup^{-1}) (w), \ \ c: = b/aa^{(1)}, \tag{4.8} \
\end{gather}
where
\begin{equation}
q^{(s)}: = \bigtriangleup ^s (q), \ s \in \mathbb{Z}. \tag{4.9}
\end{equation}
Setting
\begin{equation}
w = - \sum^\infty_{k=0} z_k \epsilon^k, \tag{4.10}
\end{equation}
we find:
\begin{equation}
(1 - \bigtriangleup^{-1}) (z_{k+1}) = (c \bigtriangleup - \bigtriangleup^{-1}c ) (1 + \bigtriangleup^{-1}) (z_k), \ 
z_0 = c, \ \ k \in \mathbb{Z}_{\geq 0}. \tag{4.11}
\end{equation}
The latter equation, being nonlocal, looks rather impenetrable; it's not even clear if it's solvable.  So let's pass 
to the continuous limit to see what the situation is in simpler circumstances.  The previous formulae become:
\begin{gather}
{a \choose b}_t = {b_x \choose ba_x} = {0 \ \ \partial b \choose b \partial \ \  0} {a \choose 1} = {0 \ \ \partial b 
\choose b \partial \ \ 0} {\delta /\delta u \choose \delta /\delta b} \bigg( \frac{a^2} {2} + b \bigg) = \tag{4.12} \\
= \left(\begin{array}{cc}
 b \partial + \partial b & a \partial b \\
 b \partial a & 2 b \partial b 
 \end{array} \right)
 {1 \choose 0} = \left(\begin{array}{cc}
 b \partial + \partial b & a \partial b \\
 b \partial a & 2 b \partial b 
 \end{array} \right)
 {\delta /\delta a \choose \delta /\delta b} (a), \notag \\
 a_t = (\epsilon b (1 - w_2))_x, \ b_t = b (a - w_1)_x, \tag{4.13} \\
 (b \partial + \partial b) (w_1) + aw_x = 0, \ w: = b w_2, \tag{4.14$a$} \\
 b \partial (aw_1 + 2 \epsilon w) = 0. \tag{4.14$b$} \
 \end{gather}
 (4.14$b$) resolves into
 \begin{gather}
 aw_1 + 2 \epsilon w = 1 \Rightarrow \tag{4.15} \\
 w_1 = (1 - 2 \epsilon w)/a \Rightarrow \notag \\
 - w_x = \frac{1}{a} (b \partial + \partial b) \frac{1}{a} (1 - 2_\epsilon w) = (\frac{b}{a^2} \partial + \partial 
\frac{b}{a^2}) (1 - 2 \epsilon w) = \notag \\
= c_x - 2 \epsilon (c \partial + \partial c) (w), \ c: = b/a^2. \tag{4.16}
\end{gather}
Setting $w = - \sum^\infty_{k=0} z_k \epsilon^k$ again (4.10), we find:
\begin{gather}
z_{k+1, x} = 2 \epsilon (c \partial + \partial c)(z_k), \ z_0 = c\Rightarrow \tag{4.17} \\
z_k = {2k+1 \choose k} c^{k+1}, \tag{4.18} \
\end{gather}
and the calculation identical to that of $\S$2 shows that (now with $\epsilon =1$)
\begin{gather}
w_1 = \eta^{-1/2}, \ \eta: = a^2 - 4b, \tag{4.19$a$} \\
w_2 = \frac{1 - a \eta^{-1/2}}{2b} . \tag{4.19$b$} \
\end{gather}
Since
\begin{equation}
\frac{\partial w_1}{\partial b} = \frac{\partial w_2}{\partial a} = 2 \eta^{-3/2}, \tag{4.20}
\end{equation}
there exists a Hamiltonian $G = G(a,b)$ such that 
\begin{equation}
{w_1 \choose w_2} = {\delta/\delta a \choose \delta/\delta b} (G), \tag{4.21}
\end{equation}
and this $G$ commutes with all the conserved densities $H_m$'s of the continuous Toda flow (4.12), because
\begin{equation}
G_{aa}/(b \partial_b)^2 (G) = \frac{1}{b} = H_{m,aa}/ (b \partial _b)^2 (H_m), \ \forall m. \tag{4.22}
\end{equation}

Thus, the continuous limit picture is manageable. Back to the discrete play, the equation (4.11.)  There exist no 
general methods to handle nonlocal recurrencies such as (4.11) save for the method of bi-Hamiltonian systems (see 
[9].)  So, we first move (4.11) into a skewsymmetric form, by applying from the left the operator (1+ 
$\bigtriangleup$), resulting in: 
\begin{equation}
(\bigtriangleup - \bigtriangleup^{-1}) (z_{k+1}) = (1 + \bigtriangleup) (c \bigtriangleup - \bigtriangleup^{-1}c) (1 + 
\bigtriangleup^{-1}) (z_k), \ z_0 = c. \tag{4.23}
\end{equation}
Unfortunately, the form (4.23), {\it{as it stands}}, is {\it{not}} of the bi-Hamiltonian character, because, e.g.,
\begin{equation}
z_1 = c^{(1)} c + cc + cc^{(1)} \not\in Im (\delta), \tag{4.24}
\end{equation}
i.e., $z_1$ is not $\delta H/\delta c$ for any $H$.  This however, shouldn't be the end of the story, and it isn't.  
The help comes from the observation that
\begin{gather}
z_1/c = \frac{\delta}{\delta c} (c^2 + c^{(-1)} c + cc^{(1)})/2, \tag{4.25$a$} \\
D (z_2/c) = [D(z_2/c)]^\dagger, \tag{4.25$b$} \
\end{gather}
so that $z_2/c \in Im (\delta)$; here $D ( \ \cdot \ )$ is the Fr$\acute{\rm{e}}$chet derivative of $ ( \ \cdot \ )$.  
This strongly suggests that we set
\begin{equation}
z_k = c \rho _k, \ k \in \mathbb{Z}_{\geq 0}, \tag{4.26}
\end{equation}
multiply eqn (4.23) from the left by $c$, and rewrite (4.23) as
\begin{equation}
[c (\bigtriangleup - \bigtriangleup^{-1})c] (\rho_{k+1}) = [c (1 + \bigtriangleup) (c \bigtriangleup - 
\bigtriangleup^{-1} c) (1 + \bigtriangleup^{-1})c] (\rho_k), \ \rho_0 =1. \tag{4.27}
\end{equation}
Miraculously, and for no discernible reason:
\\
(a) \ The matrix (in fact, scalar)
\begin{equation}
b^2 = c(1 + \bigtriangleup) (c \bigtriangleup - \bigtriangleup^{-1} c) (1 + \bigtriangleup^{-1} )c \tag{4.28}
\end{equation}
is Hamiltonian;
\\
(b) The pair of Hamiltonian matrices, $b^2$ (4.28) and 
\begin{equation}
b^1 = c (\bigtriangleup - \bigtriangleup^{-1}) c \tag{4.29}
\end{equation}
form a Hamiltonian pair.  The bi-Hamiltonian theory then guarantees the existence of a sequence of Hamiltonians 
$\{h_m\}$ such that
\begin{equation}
\rho_m = \delta h_m/\delta c, \ \ \ m \in \mathbb{Z}_{\geq 0}. \tag{4.30}
\end{equation}
Thus, our constrain (4.8) has been resolved:
\begin{equation}
w = bw_2 = - \sum^\infty_{k=0} \epsilon^k c (\delta h_k/\delta c), \ \ c = b/aa^{(1)}. \tag{4.31}
\end{equation}
Setting
\begin{equation}
h = \sum^\infty_{k=0} \epsilon^k h_k, \tag{4.32}
\end{equation}
and noticing that
\begin{equation}
\frac{\delta}{\delta a} = - \frac{1}{a} (1 + \bigtriangleup^{-1}) c \frac{\delta }{\delta c}, \ \ \frac{\delta}{\delta 
b} = \frac{1}{aa^{(1)}} \frac{\delta}{\delta c}, \tag{4.33}
\end{equation}
we see that
\begin{gather}
-w_2 = \frac{1}{b} \sum^\infty_{k=0} c \epsilon^k \frac{\delta h_k}{\delta c} =  \frac{1}{a \ a^{(1)}} \frac{\delta 
h}{\delta c} = \frac{\delta h}{\delta b}, \tag{4.34$a$} \\
w_1 = \frac{1}{a} - \epsilon \frac{1}{a} (1 + \bigtriangleup^{-1}) (- c \frac{\delta h}{\delta c}) = \frac{1}{a} - 
\frac{\delta h}{\delta a} = \frac{\delta}{\delta a} (log \ a-h). \tag{4.34$b$} \
\end{gather}
Thus,
\begin{equation}
{w_1 \choose w_2} = {\delta /\delta a \choose \delta /\delta b} (log \ a-h), \tag{4.35}
\end{equation}
and the last step now is to show that $log a-h$ commutes with all the conserved densities $H_m$'s of the full Toda 
lattice (4.3).  But this is true {\it{in general}}:
\\
$(\hat C)$ Suppose that the constrain $B^2 (w) = 0$ is (1.5) has been resolved as
\begin{equation}
w = \frac{\delta G}{\delta u}, \ \ {\rm{some}} \ G. \tag{4.36}
\end{equation}
Then
\begin{gather}
\{G, H_m\} \sim \frac{\delta H_m}{\delta u^t} B^1 (\frac{\delta G}{\delta u}) \sim - \bigg[B^1 \bigg(\frac{\delta 
H_m}{\delta u}\bigg)\bigg]^t \frac{\delta G}{\delta u} = \notag \\
= - \bigg[B^2 \bigg(\frac{\delta H_{m-1}}{\delta u}\bigg)\bigg]^t \frac{\delta G}{\delta u} \sim \frac{\delta 
H_{m-1}}{\delta u^t} B^2 \bigg(\frac{\delta G}{\delta u}\bigg) =  \frac{\delta H_{m-1}}{\delta u^t} B^2 (w) = 0. 
\tag{4.37}
\end{gather}
Thus, our prescription for analyzing the nonholonomic deformation, stated at the end of Section 2, works perfectly 
provided the nonholonomic perturbation $w$ is a variational derivative (4.36).
\\
{\bf{Remark 4.38.}} With hindsight, one readily sees that the bi-Hamiltonian pair $b^1$ (4.29) and $b^2$ (4.28) 
describes the classical Volterra lattice
\begin{equation}
\dot c = c (c^{(1)} - c^{(-1)}), \tag{4.39}
\end{equation}
see [7].
\\
{\bf{Remark 4.40.}} The nonholonomic perturbation (1.5) of the Volterra lattice (4.39),
\begin{equation}
\dot c = c (\bigtriangleup - \bigtriangleup^{-1}) c (1 - w), \ \ \ (c \bigtriangleup - \bigtriangleup^{-1} c) (1 + 
\bigtriangleup^{-1}) (w) = 0, \tag{4.41}
\end{equation}
is not simplified by the rescaling $c \mapsto \epsilon c$, but it does so upon the rescaling $c \mapsto 1 + \epsilon 
c$ around the {\it{stationary}} solution $\{ c = 1\}$ of the Volterra lattice.

\section{The Euler Top}

The constrain $B^2 (w) = 0$ is, in general, nonholonomic only for systems which are either differential or difference 
on $\mathbb{Z}$, i.e., for dimensions 1 and ``1/2".  In $0$ dimensions, i.e., in Classical Mechanics with a 
{\it{finite}} number of degrees of freedom, the constrain $B^2 (w) = 0$ becomes {\it{holonomic.}}  Thus, e.g., there 
would be no problem in resolving that constrain for the {\it{periodic}} Toda lattice on $\mathbb{Z}/N\mathbb{Z}$.  

Let's see how this works in practice, for the simplest possible case, the so(3) Euler top:
\begin{equation}
\dot x_1 = \alpha_1 x_2 x_3, \ \ \dot x_2 = \alpha_2 x_3 x_1, \ \ \dot x_3 = \alpha_3 x_1 x_2, \tag{5.1}
\end{equation}
where $x = (x_1, x_2, x_3) \in \mathbb{R}^3.$  
I follow the notation and results of a short and concise presentation in [12].  Let's agree that $(ijk)$ stands for an 
even permutation of (123).  Equations (5.1) can be the rewritten as
\begin{equation}
\dot x_i = \alpha_i x_j x_k, \tag{5.2}
\end{equation}
with $\mbox{\boldmath$\alpha$} = (\alpha_1, \alpha_2, \alpha_3) \in \mathbb{R}^3$ being an arbitrary but fixed vector 
of parameters.  If ${\bf{c}}= (c_1, c_2, c_3)$ is another vector in $\mathbb{R}^3$, then
\begin{equation}
H_c = \frac{1}{2} \sum^3_{i=1} c_i x_i^2 \tag{5.3}
\end{equation}
is an integral of (5.1) iff
\begin{equation}
(\alpha, {\bf{c}}) = 0, \tag{5.4}
\end{equation}
so that the space of integrals is two-dimensional.  

Now, for any vector $\mbox{\boldmath$\gamma$} \in \mathbb{R}^3$, consider the Poisson brackets
\begin{equation}
\{x_i, x_j\}^{(\gamma)} = \gamma_k x_k. \tag{5.5}
\end{equation}
Then
\begin{equation}
\{x_i, H_c\}^{(\gamma)} = c_j x_j \gamma_k x_k - c_k x_k \gamma_j x_j = (c_j \gamma_k - c_k \gamma_j) x_j x_k, 
\tag{5.6}
\end{equation}
so that the conditions
\begin{equation}
\mbox{\boldmath$\alpha$} = {\bf{c}} \times \mbox{\boldmath$\gamma$} \ \Rightarrow (\mbox{\boldmath$\alpha$}, {\bf{c}}) 
= 0, \tag{5.7}
\end{equation}
guarantee that the motion equations (5.1) are Hamiltonian in the Hamiltonian structure (5.5) with the Hamiltonian 
(5.3).  So, let's choose vectors $\mbox{\boldmath$\beta$}$ and $\mbox{\boldmath$\gamma$}$ such that 
$\{\mbox{\boldmath$\beta$}, \mbox{\boldmath$\gamma$}, \mbox{\boldmath$\alpha$}/|\mbox{\boldmath$\alpha$}|\}$ from a 
right orthonormal basis.  Set $B^1$ and $B^2$ corresponding to $\mbox{\boldmath$\beta$}$ and 
$\mbox{\boldmath$\gamma$}$, respectively:
\begin{equation}
-B^1 = \left(\begin{array}{ccc}
 0 & - \beta_3 x_3 & -\beta_2 x_2 \\
\beta_3 x_3 & 0 & \beta_1 x_1 \\
\beta_2 x_2 & - \beta_1 x_1 & 0  
 \end{array} \right), 
 - B^2 = \left(\begin{array}{ccc}
 0 & - \gamma_3 x_3 & -\gamma_2 x_2 \\
\gamma_3 x_3 & 0 & \gamma_1 x_1 \\
\gamma_2 x_2 & - \gamma_1 x_1 & 0  
 \end{array} \right). \tag{5.8}
 \end{equation}
 The constrain $B^2 (w) = 0$ becomes:
 \begin{equation}
 {\bf{w}} \times \mathbb{X}^{(\gamma)} = 0, \ \ \mathbb{X}^{(\gamma)} : = (\gamma_1 x_1, \gamma_2 x_3, \gamma_3  x_3), 
\tag{5.9}
\end{equation}
so that
\begin{equation}
{\bf{w}}= const \mathbb{X}^{(\gamma)}, \ \ const \ = const (t), \tag{5.10}
\end{equation}
and the perturbed motion equations (1.5) become:
\begin{equation}
\dot x_i = \alpha_i x_j x_k - const (\mathbb{X}^{(\beta)} \times \mathbb{X}^{(\gamma)})_i. \tag{5.11}
\end{equation}
But
\begin{equation}
(\mathbb{X}^{(\beta)} \times \mathbb{X}^{(\gamma)})_i = (\mbox{\boldmath$\beta$} \times \mbox{\boldmath$\gamma$})_i 
x_j x_k = \frac{1}{|\alpha|} \alpha_i x_j x_k \tag{5.12}
\end{equation}
Thus, finally, the perturbed top equations are
\begin{equation}
\dot x_i = const^\prime \alpha_i x_j x_k,  \tag{5.13}
\end{equation}
so that the overall effect of the perturbation amounts to the time rescaling of the original top.  This is reminiscent 
of the general Chaplygin theorem identifying some special nonholonomic systems with the time-rescaled Hamiltonian ones 
(see [4,3].)

\end{document}